\documentstyle[psfig]{mn}

\newif\ifAMStwofonts

\ifoldfss
  \ifCUPmtlplainloaded \else
    \NewTextAlphabet{textbfit} {cmbxti10} {}
    \NewTextAlphabet{textbfss} {cmssbx10} {}
    \NewMathAlphabet{mathbfit} {cmbxti10} {} 
    \NewMathAlphabet{mathbfss} {cmssbx10} {} 
  \fi
  \ifAMStwofonts
    \ifCUPmtlplainloaded \else
      \NewSymbolFont{upmath} {eurm10}
      \NewSymbolFont{AMSa} {msam10}
      \NewMathSymbol{\upi}     {0}{upmath}{19}
      \NewMathSymbol{\umu}     {0}{upmath}{16}
      \NewMathSymbol{\upartial}{0}{upmath}{40}
      \NewMathSymbol{\leqslant}{3}{AMSa}{36}
      \NewMathSymbol{\geqslant}{3}{AMSa}{3E}

      \let\leq=\leqslant 
       
    \fi
  \fi
\fi 

\ifnfssone
  \newmathalphabet{\mathit}
  \addtoversion{normal}{\mathit}{cmr}{m}{it}
  \addtoversion{bold}{\mathit}{cmr}{bx}{it}
  \newmathalphabet{\mathbfit} 
  \addtoversion{normal}{\mathbfit}{cmr}{bx}{it}
  \addtoversion{bold}{\mathbfit}{cmr}{bx}{it}
  \newmathalphabet{\mathbfss} 
  \addtoversion{normal}{\mathbfss}{cmss}{bx}{n}
  \addtoversion{bold}{\mathbfss}{cmss}{bx}{n}
  \ifAMStwofonts
    \ifCUPmtlplainloaded \else
      \UseAMStwoboldmath
      \makeatletter
      \new@mathgroup\upmath@group
      \define@mathgroup\mv@normal\upmath@group{eur}{m}{n}
      \define@mathgroup\mv@bold\upmath@group{eur}{b}{n}
      \edef\UPM{\hexnumber\upmath@group}
      \new@mathgroup\amsa@group
      \define@mathgroup\mv@normal\amsa@group{msa}{m}{n}
      \define@mathgroup\mv@bold\amsa@group{msa}{m}{n}
      \edef\AMSa{\hexnumber\amsa@group}
      \makeatother
      \mathchardef\upi="0\UPM19
      \mathchardef\umu="0\UPM16
      \mathchardef\upartial="0\UPM40
      \mathchardef\leqslant="3\AMSa36
      \mathchardef\geqslant="3\AMSa3E

      \let\leq=\leqslant 

    \fi
  \fi
\fi 

\ifnfsstwo
  \DeclareMathAlphabet{\mathbfit}{OT1}{cmr}{bx}{it}
  \SetMathAlphabet\mathbfit{bold}{OT1}{cmr}{bx}{it}
  \DeclareMathAlphabet{\mathbfss}{OT1}{cmss}{bx}{n}
  \SetMathAlphabet\mathbfss{bold}{OT1}{cmss}{bx}{n}
  \ifAMStwofonts
    \ifCUPmtlplainloaded \else
      \DeclareSymbolFont{UPM}{U}{eur}{m}{n}
      \SetSymbolFont{UPM}{bold}{U}{eur}{b}{n}
      \DeclareSymbolFont{AMSa}{U}{msa}{m}{n}
      \DeclareMathSymbol{\upi}{0}{UPM}{"19}
      \DeclareMathSymbol{\umu}{0}{UPM}{"16}
      \DeclareMathSymbol{\upartial}{0}{UPM}{"40}
      \DeclareMathSymbol{\leqslant}{3}{AMSa}{"36}
      \DeclareMathSymbol{\geqslant}{3}{AMSa}{"3E}

      \let\leq=\leqslant 

    \fi
  \fi
\fi 

\ifCUPmtlplainloaded \else
  \ifAMStwofonts \else 
    \def\upi{\pi}
    \def\umu{\mu}
    \def\upartial{\partial}
  \fi
\fi

\title{A Local Void and the Accelerating Universe}
\author[K. Tomita]
       {K. Tomita \\
        Yukawa Institute for Theoretical Physics, Kyoto University,
        Kyoto, 606-8502, Japan}

\pagerange{\pageref{firstpage}--\pageref{lastpage}}
\pubyear{2000}
\newlength{\plotwidth}
\begin{document}

\maketitle

\label{firstpage}

\begin{abstract}
Corresponding to the recent observational claims that we are in a
local void (an underdense region) on scales of $200  - 300$ Mpc, 
the magnitude-redshift
relation in a cosmological model with a local void is investigated. It 
is already evident that the accelerating behavior of high-z supernovae 
can be explained in this model, because the local void plays a role
similar to the positive cosmological constant. In this paper the
dependence of the behavior on the gaps of cosmological parameters in
the inner (low-density) region and the outer (high-density) region, the
radius of the local void, and the clumpiness parameter is studied
and its implication is discussed.  
\end{abstract}


\section{Introduction}

One of the most important cosmological observations at present is the
[$m, z$] relation for high-z supernovae (SNIa), which play a role of
standard candles at the stage reaching epochs $z \ga 1$. 
So far the observed data of SNIa have been compared with the
theoretical relation in homogeneous and isotropic models, and many
workers have made efforts to determine their model parameters
(Garnavich et al. 1998, Schmidt et al. 1998, Perlmutter et al. 1999,
Riess et al. 1998, 2000, Riess 2000). 
 
Here is, however, an essentially important problem to be taken
into consideration. It is the homogeneity of the Universe. According to
Giovanelli et al. (1998, 1999) and Dale et al. (1999)
the Universe is homogeneous in the region within $70 h^{-1}$ Mpc \
(the Hubble constant $H_0$ is $100 h$ km s$^{-1}$ Mpc$^{-1}$). 
On the other hand,
recent galactic redshift surveys (Marinoni et al. 1999, Marzke et al. 1998,
Folkes et al. 1999, Zucca et al. 1997) show that in the region around 
$200 - 300 h^{-1}$ Mpc from us the distribution of galaxies may be 
inhomogeneous. This is because the galactic 
number density in the region of $z < 0.1$ or $< 300 h^{-1}$ Mpc from
us was shown to be by a factor $> 1.5$ smaller than that in the
remote region of $z > 0.1$. Recently a large-scale inhomogeneity
suggesting a wall around the void on scales of $\sim 250 h^{-1}$ Mpc
has been found by Blanton et al. (2000) in the SDSS commissioning data 
(cf. their Figs. 7 and 8). Similar walls on scales of $\sim 250 h^{-1}$ Mpc
have already been found in the Las Campanas and 2dF redshift surveys
near the Northern and Southern Galactic Caps (Shectman et al. 1996, 
Folkes et al. 1999, Cole et al. 2000).
These results mean that there is a local void with the radius of 
$200 - 300 h^{-1}$ Mpc and we live in it. 
 
Moreover, the measurements by Hudson et al. (1999) and Willick (1999) for a
systematic deviation of clusters' motions from the global Hubble flow
may show some inhomogeneity on scales larger than $100 h^{-1}$ Mpc.
Another suggestion for inhomogeneity comes from the
periodic wall structures on scales of $\sim 130 h^{-1}$ Mpc, as have
been shown by Broadhurst et al. (1990), Landy et al. (1996), and 
Einasto et al. (1997). This is connected with the anomaly of the power
spectrum around $100 - 200 h^{-1}$ Mpc (so-called ``excess power'')
which was discussed by Einasto et al. (1999). 
This fact also may suggest some inhomogeneity in the above nearby region.

If the local void exists really, the Hubble constants also must be
inhomogeneous, as well as the density parameters, and the theoretical
relations between observed quantities are different from those in
homogeneous models. At present, however, the large-scale inhomogeneity 
of the Hubble constant has not been observationally established yet
because of the large error bars in the various measurements (cf. Tomita 2001).

In my previous papers (Tomita 2000a and 2000b) cited as Paper 1 and Paper 
2, I showed various models with a local void and discussed the bulk
flow, CMB dipole anisotropy, distances and the [$m, z$] relation in
them in the limited parameter range. It was found that the
accelerating behavior of supernovae can be explained in these models
without cosmological constant. On the other hand, Kim et al. (1997)
showed that the difference between the local and global values of the
Hubble constant should be smaller than $10 \%$ in {\it homogeneous} 
cosmological models in order to be consistent with the SNIa data. 
However, this does not impose any strong condition on the
difference in inhomogeneous models, because their analyses were done
using the luminosity distance in {\it homogeneous} models and so
they are incomplete.  In fact my previous papers showed 
concretely that in inhomogeneous models larger differences can be 
consistent with the data. The possibility that the above difference 
may explain the behavior of SNIa was later discussed also by Goodwin 
et al. (1999).  
  
In this paper I describe first (in \S 2) a simplified cosmological
model with a local void and treat distances in light paths with
nonzero clumpiness (smoothness) parameter $\alpha$. In the previous
paper (Paper 2), I considered only distances in full-beam light paths 
($\alpha = 1$), but in the realistic paths there are deviations from 
$\alpha = 1$ due to lensing effects from inhomogeneous matter distributions. 
In \S 3, I show the
dependence of the [$m, z$] relation on model parameters such as the radius
of the local void, the ratios of density parameters and Hubble
constants in the inner (low-density) and outer (high-density) regions, and
the clumpiness parameter. The constraints to the parameters are
derived in comparison between the above relations in the present
models and the relations in homogeneous models.
 
Finally (in \S 4), we discuss the remaining problems and describe
concluding remarks.

\section[]{Distances in models with a local void}

The inhomogeneous models we consider consist of inner (low-density) region 
V$^{\rm I}$ and outer (high-density) region V$^{\rm II}$ which are
separated by a single shell. It is treated as a spherical singular
shell and the mass in it compensates the mass deficiency in V$^{\rm
I}$. So V$^{\rm I}$ and the shell are regarded as a local void and the 
wall, respectively. The line-element in the two regions are
\begin{eqnarray}
  \label{eq:m1}
 ds^2 &=& g^j_{\mu\nu} (dx^j)^\mu (dx^j)^\nu \cr 
&=& - c^2 (dt^j)^2 + [a^j
 (t^j)]^2 \Big\{ d (\chi^j)^2 + [f^j (\chi^j)]^2 d\Omega^2 \Big\},
\end{eqnarray}
where $j \ (=$ I or II) represents the regions, $f^j (\chi^j) = \sin
\chi^j, \chi^j$ and $\sinh \chi^j$ for $k^j = 1, 0, -1$, respectively,
and $d\Omega^2 = d\theta^2 + \sin^2 \theta  d\varphi^2$. 
In the following the negative curvature is assumed in all regions. 
The Hubble constants 
and density parameters are expressed as $(H_0^{\rm I}, H_0^{\rm II})$ and
$(\Omega_0^{\rm I}, \Omega_0^{\rm II})$, where we assume that
$H_0^{\rm I} > H_0^{\rm II}$ and $\Omega_0^{\rm I} < \Omega_0^{\rm II}$. 
 The distances of the shell and the observer O (in V$^{\rm I}$) from 
the centre C (in V$^{\rm I}$) are assumed to be
$200$ and $40 h_{\rm I}^{-1}$ as a standard case.
 This shell corresponds to the redshift $\bar{z}_1 = 0.067$ \
(see Figure 1).

\begin{figure}
\centerline{\psfig{figure=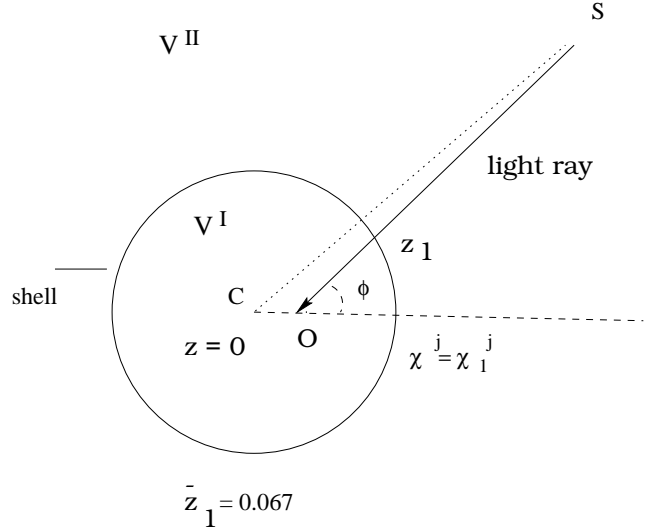,width=\plotwidth}}
\caption{Model with a spherical single shell. Redshifts for observers 
at O and C are $z$ and $\bar{z}$.
  \label{fig:1}}
\end{figure}

In Paper 2 we derived the full-beam distances ($\overline{\rm CS}$) between the
centre C and a source S, and the distances ($\overline{\rm OS}$) between an
observer O and S. The two distances are nearly equal in the case when 
$\overline{\rm CS}$ or $\overline{\rm OS}$ is much larger than 
$\overline{\rm CO}$. Since we notice
this case alone in the following, we treat the light paths $\overline{\rm CS}$
for simplicity. Then the angular-diameter distance $d_{\rm A}$ is
\begin{equation}
  \label{eq:m2}
d_{\rm A} = a^{\rm I}(\bar{\eta}^{\rm I}_{\rm s}) \sinh
(\bar{\chi}^{\rm I}_{\rm s}), 
\end{equation}
if a source S is in V$^{\rm I}$, where $(\bar{\eta}^{\rm I}_{\rm s},
\bar{\chi}^{\rm I}_{\rm s})$ are the coordinates of S, and $\eta$ is
the conformal time coordinate. Here bars are used for the coordinates 
along the light paths to
the virtual observer at C. If S is in V$^{\rm II}$, we
have
\begin{eqnarray}
  \label{eq:m3}
d_{\rm A} = a^{\rm I}(\bar{\eta}^{\rm I}_1) \sinh (\bar{\chi}^{\rm I}_1) 
&+& [a^{\rm II}(\bar{\eta}^{\rm II}_{\rm s} \sinh (\bar{\chi}^{\rm II}_{\rm
s})\cr
 &-& a^{\rm II}(\bar{\eta}^{\rm II}_1) \sinh (\bar{\chi}^{\rm II}_1)],
\end{eqnarray}
where $(\bar{\eta}^{\rm I}_{\rm 1}, \bar{\chi}^{\rm I}_{\rm 1})$ stand 
for the shell, and we have
\begin{equation}
  \label{eq:m4}
a^{\rm I}(\bar{\eta}^{\rm I}_1) \sinh (\bar{\chi}^{\rm I}_1) =
a^{\rm II}(\bar{\eta}^{\rm II}_1) \sinh (\bar{\chi}^{\rm II}_1)
\end{equation}
from the junction condition. 

Here we treat the following equation for
the angular-diameter distance to consider the clumpiness along paths
into account (Dyer and Roeder 1973, Schneider et al. 1992, Kantowski
1998, Tomita 1999):
\begin{eqnarray}
  \label{eq:m5}
{d^2 (d_{\rm A}^j) \over d(z^j)^2} &+& \Big\{{2 \over 1+z^j}+{1 \over
2}(1+z^j) \Big[\Omega_0^j (1+3z^j) 
+ 2 -2\lambda_0^j\Big]\cr
&& \times F^{-1} \Big\}{d (d_{\rm A}^j) \over dz^j} 
+ {3 \over 2} \Omega_0^j \alpha (1+z^j)
F^{-1} d_{\rm A}^j = 0,
\end{eqnarray}
where
$j = $ I and II, $z^j$ is the redshift in the region V$^j$, $\alpha$
is the clumpiness parameter, and
\begin{equation}
  \label{eq:m6}
F \equiv (1+ \Omega_0^j z^j)(1+z^j)^2 - \lambda_0^j z^j (2+ z^j).
\end{equation}
Here and in the following the bars are omitted for simplicity.
The two redshifts at the shell are equal, i.e.
\begin{equation}
  \label{eq:m7}
 z_1^{\rm I} = z_1^{\rm II} \ (\equiv z_1)
\end{equation}
for the comoving shell (cf. Paper I).

The distances $d_{\rm A}^{\rm I}$ in V$^{\rm I}$ is obtained solving
Eq.(\ref{eq:m5}) under the conditions at $z^{\rm I} = 0$:
\begin{equation}
  \label{eq:m8}
(d_{\rm A}^{\rm I})_0 = 0, \quad (d_{\rm A}^{\rm I}/dz^{\rm I})_0 =
c/H_0^{\rm I}, 
\end{equation}
and $d_{\rm A}^{\rm II}$ in V$^{\rm II}$ is obtained similarly under
the conditions at $z^{\rm II} = 0$: 
\begin{equation}
  \label{eq:m9}
(d_{\rm A}^{\rm II})_0 = const, \quad (d_{\rm A}^{\rm II}/dz^{\rm II})_0 =
c/H_0^{\rm II}, 
\end{equation}
where $const$ is determined so that the junction condition $d_{\rm
A}^{\rm I}(z_1) = d_{\rm A}^{\rm II}(z_1)$ may be satisfied at the shell.
Then the distance $d_{\rm A} (z_s)$ from C to the source S is
\begin{equation}
  \label{eq:m10}
d_{\rm A} (z_s) = d_{\rm A}^{\rm I}(z_s) \hspace{3cm} {\rm for} \ z_s \leq z_1,
\end{equation}
and
\begin{equation}
  \label{eq:m11}
d_{\rm A} (z_s) = d_{\rm A}^{\rm I}(z_1) + d_{\rm A}^{\rm II}(z_s)
 - d_{\rm A}^{\rm II}(z_1) \quad {\rm for} \ z_s > z_1,
\end{equation}
where $z_s = z_s^{\rm I}$ and $z_s^{\rm II}$ for $z_s \leq z_1$ and
$z_s > z_1$, respectively. The luminosity distance $d_{\rm L}$ is 
related to the angular-diameter distance $d_{\rm A}$ \ by 
\ $d_{\rm L} = (1 + z)^2 d_{\rm A}$.

As for the clumpiness parameter $\alpha$, we studied the distribution
function $N(\alpha)$ as a function of $z$ in our previous
papers (Tomita 1998, Tomita 1999).  To obtain  $N(\alpha)$,
we first derived model
universes consisting of galaxies and halos using a N-body simulation
technique, secondly calculated the angular-diameter distance by
solving null-geodesic equations along many light paths between an
observer and sources at epoch $z$, and finally derived a statistical
distribution of $\alpha$ determined in a comparison with the Friedmann 
distance ($\alpha = 1.0$) and the Dyer-Roeder distance ($\alpha = 0.0$). 
As the result of these studies, it was found 
that the average value $\bar{\alpha}$ of $\alpha$ is $1.0$, which
represents the Friedmann distance, and the dispersion $\sigma_\alpha$
can be $\sim 0.5$ for $z <2.0$. If the detection of high-z supernovae
is done in completely random directions, the observed average value
of  $\alpha$ is equal to the above theoretical average value
$\bar{\alpha}$. But if the detections are biased to the directions
with less galactic number per steradian to avoid the dust obscuration, 
we may have the value of $\alpha \sim \bar{\alpha} -
\sigma_\alpha$. Then the angular-diameter and luminosity distances are 
somewhat longer than the average Friedmann distances. In the next
section we show the cases with $\alpha = 1.0, 0.5,$ and $0.0$ for
comparison.
The lensing effect on the [$m, z$] relation of SNIa was discussed also 
by Holz (1998), Porciani and Madau (2000), and Barber (2000).

\begin{figure}
\centerline{\psfig{figure=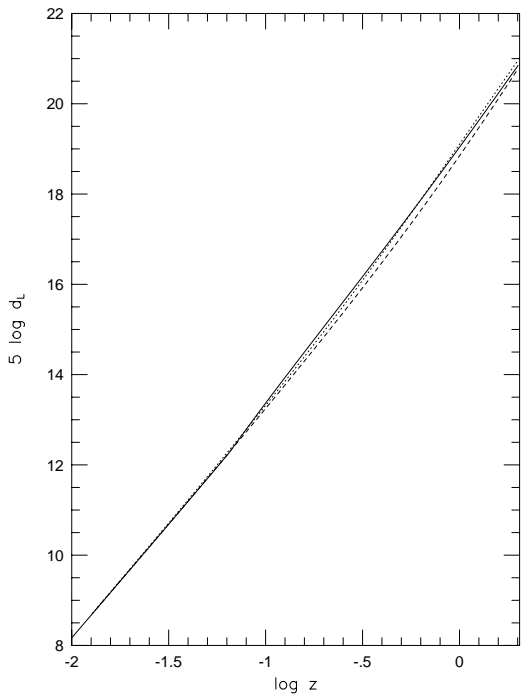,width=\plotwidth}} 
\caption{The [$m, z$] relation in cosmological models with a local
void. The solid line denotes the case with a standard parameter set
given in (\ref{eq:p1}). The dotted and dash lines stand for
homogeneous models with $(\Omega_0, \lambda_0) = (0.3, 0.7)$ and $(0.3,
0.0)$, respectively, for comparison.
  \label{fig:2}}
\end{figure}

\section{Parameter dependence of the magnitude-redshift relation}

As for homogeneous models it is well-known from the comparison with 
observational data that the flat case with
nonzero cosmological constant of $(\Omega_0, \lambda_0) = (0.3, 0.7)$
can represent the accelerating behavior of high-z SNIa, while an open
model with $(0.3, 0)$ cannot explain their data for $z \approx 1.0$ 
(Garnavich, et al. 1998, Schmidt et al. 1998, Perlmutter et al. 1999,
Riess et al. 1998, 2000, Riess 2000).  In the present inhomogeneous
models we have six model parameters $(\Omega_0^{\rm I}, 
\lambda_0^{\rm I}, H_0^{\rm I}, H_0^{\rm II}/H_0^{\rm I}, \Omega_0^{\rm II}
/\Omega_0^{\rm I}, z_1)$, and their direct fitting with the observational 
data is much complicated in contrast to the homogeneous case with
three parameters. In this paper the parameter dependence of [$m,
z$] relations are examined for the preliminary study and the relations 
in these two homogeneous models are used as a measure
for inferring how the relations in inhomogeous models  with various
parameters can reproduce the observational data. That is, we deduce
that the model parameters are consistent with the observational data, 
if at the interval $0.5 < z < 1.0$ the curve in the [$m, z$] relation 
is near that in the homogeneous
model $(0.3, 0.7)$ comparing with the difference between those for
$(0.3, 0.7)$ and $(0.3, 0)$. 

For the [$m, z$] relation in an inhomogeneous model, we first treat the 
case with the following {\it standard} parameters to reproduce the 
accelerating 
behavior in the above homogeneous model $(0.3, 0.7)$ in a similar way:
\begin{eqnarray}
  \label{eq:p1}
(\Omega_0^{\rm I}, \Omega_0^{\rm II}) &=& (0.3, 0.6),\cr 
H_0^{\rm I} &=& 71, \quad H_0^{\rm II}/H_0^{\rm I} = 0.82, \cr
\alpha &=& 1.0,  \quad z_1 = 0.067, \ {\rm and} \ \lambda_0^{\rm I} =
\lambda_0^{\rm II} = 0.
\end{eqnarray}
The radius of the local void is $r_1 \equiv (c/H_0^{\rm I}) z_1 = 200
(h_{\rm I})^{-1}$ Mpc. In Figure 2, the relation is shown for $z =
0.01 - 2.0$ in comparison with that in two homogeneous models with
parameters: $(\Omega_0, \lambda_0) = (0.3, 0.7)$ and $(0.3, 0), H_0 =
71$ and $\alpha = 1.0$. For $z<z_1$ the relation is equal to that in
the open model $(0.3, 0.0)$.

It is found that the behavior in the case of 
$(\Omega_0^{\rm I}, \Omega_0^{\rm II}) = (0.3, 0.6)$ with $\lambda_0^{\rm I} =
\lambda_0^{\rm II} = 0$ accords approximately with that in the flat,
homogeneous model with $(\Omega_0, \lambda_0) = (0.3, 0.7)$ for $z_1 < z 
< 1.0$. Accordingly they is similarly fit for the observed data of SNIa.

Next, to examine the parameter dependence of the [$m, z$] relation, we 
take up various cases with following parameters different from the 
above standard case:
\begin{eqnarray}
  \label{eq:p2}
\Omega_0^{\rm II} &=& 0.45 \ {\rm and} \ 0.80 \quad ({\rm for} \
\Omega_0^{\rm I} = 0.3)\cr
H_0^{\rm II}/H_0^{\rm I} &=& 0.80 \ {\rm and}\ 0.87 \quad ({\rm for} \ 
 H_0^{\rm I} = 71), \cr
\alpha &=& 0.0 \ {\rm and} \ 0.5,  \quad z_1 = 0.05 \ {\rm and} \ 
0.167, \ {\rm and} \cr
 \lambda_0^{\rm II} &=& 0.4.
\end{eqnarray}
Here, for $z_1 = 0.05$ and $0.167$, we have $r_1 = (150$ and
$500)(h_{\rm I})^{-1}$ Mpc, respectively. Since $\lambda_0^j \equiv {1 
\over 3} \Lambda (c/H_0^j)^2,$ \ we have $\lambda_0^{\rm I} =
\lambda_0^{\rm II} (H_0^{\rm II}/H_0^{\rm I})^2.$

\begin{figure}
\centerline{\psfig{figure=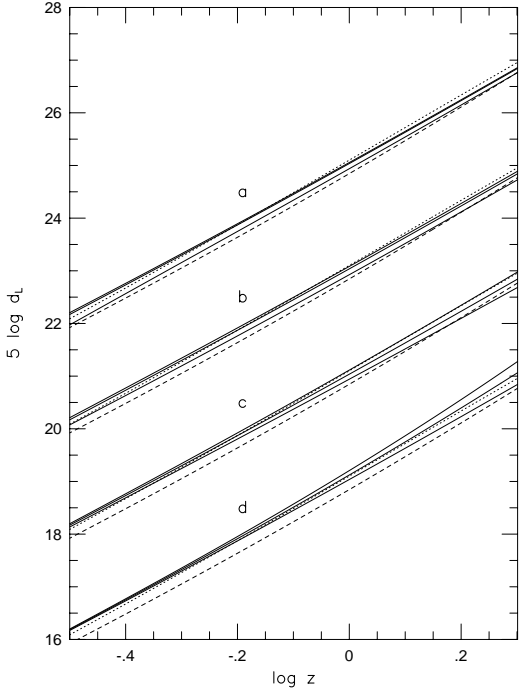,width=\plotwidth}}    
\caption{The [$m, z$] relation in cosmological models with a local
void. The solid lines denote (a) the cases with $z_1 = 0.05, 0.067,$ and
$0.167$ (from the top to the bottom), which correspond to the shell 
radius $r_1 = 150, 200$, and $500 (h_{\rm I})^{-1}$ Mpc, 
respectively, (b) the cases with 
$H_0^{\rm II}/H_0^{\rm I} = 0.87, 0.82,$ and $0.80$ from the top 
to the bottom, (c) the cases with 
$\Omega_0^{\rm II} = 0.45, 0.6,$ and $0.8$ from the top to the bottom,
and (d) the cases with $\alpha = 0.0, 0.5,$ and
$1.0$ from the top to the bottom.
The other parameters are same as
those in a standard parameter set given in (\ref{eq:p1}). The dotted
and dash lines stand for homogeneous models, as in Figure 2.
Curves (a), (b) and (c) were depicted in the single figure together
with (d) by shifting upward as $\Delta (5 \log d_{\rm L}) = 6, 4,
2$, respectively.  
  \label{fig:3}}
\end{figure}

\begin{figure}
\centerline{\psfig{figure=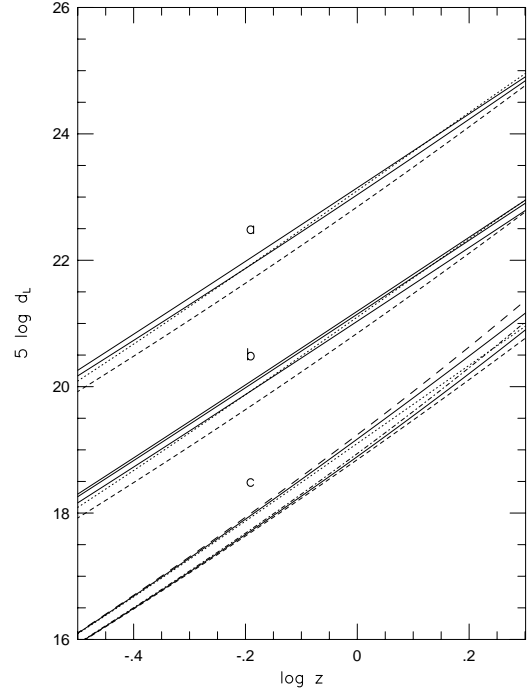,width=\plotwidth}} 
\caption{The [$m, z$] relation in cosmological models with a local
void. The upper and lower solid lines denote (a) the cases with 
$\lambda_0^{\rm II} = 0.4$ and $0.0$, respectively, and  
(b) the cases with $H_0^{\rm II}/H_0^{\rm I} =
0.87, 0.82,$ and $0.80$ for $\lambda_0^{\rm II} = 0.4$ from the top to
the bottom. (c) The [$m, z$] relation in two homogeneous 
cosmological models
with  $(\Omega_0, \lambda_0) = (0.3, 0.7)$ and $(0.3, 0.0)$.
The upper and lower groups of three lines stand for models $(0.3,
0.7)$ and $(0.3, 0.0)$. The upper, middle and lower lines in each
group are for $\alpha = 0.0, 0.5,$ and $1.0$, respectively. 
The other parameters are same as
those in a standard parameter set given in (\ref{eq:p1}). The dotted
and dash lines stand for homogeneous models, as in Figure 2.
Curves (a) and (b) were  depicted in the single figure together
with (c) by shifting upward as $\Delta (5 \log d_{\rm L}) = 4,
2$, respectively.  
  \label{fig:4}}
\end{figure}

In Figure 3 (curves a), the cases with $z_1 = 0.05, 0.067,$ and 
$0.167$ are shown 
in a model with $(\Omega_0^{\rm I}, \Omega_0^{\rm II}) = (0.3, 0.6), 
\ H_0^{\rm II}/H_0^{\rm I} = 0.82, \ \alpha = 1.0, \ {\rm and} \ 
\lambda_0^{\rm I} = \lambda_0^{\rm II} = 0.$  The range of $z$ was
changed to $0.3 < z < 2.0$ to magnify the figures.
From this Figure it is
found that if $r_1 = 150$ and $200  (h_{\rm I})^{-1}$ Mpc, the [$m,
z$] relation is similar to that in the flat, homogeneous model with 
$(\Omega_0, \lambda_0) = (0.3, 0.7)$ for $z \sim 0.5$, but if  $r_1 =
500  (h_{\rm I})^{-1}$ Mpc, the relation is rather different from that 
in the latter model. This means that $r_1$ must be $\la 300 (h_{\rm
I})^{-1}$ Mpc to explain  the [$m,z$] relation of SNIa. This
observational constraint is consistent with the observationally
estimated radius of the local void ($\la 300 h^{-1}$ Mpc) (cf. 
Marinoni 1999, Marzke et al. 1998, Folkes et al. 1999, Zucca 
et al. 1997).

In Figure 3 (curves b), the cases with $H_0^{\rm II}/H_0^{\rm I} 
= 0.80, 0.82,$
and $0.87$ are shown in a model with  $(\Omega_0^{\rm I}, 
\Omega_0^{\rm II}) = (0.3, 0.6), \ z_1 = 0.067, \ \alpha = 1.0, \ 
{\rm and} \ \lambda_0^{\rm I} = \lambda_0^{\rm II} = 0.$
In the cases with $H_0^{\rm II}/H_0^{\rm I} = 0.82, 0.87,$ the
relations are found to be consistent with the relation in the above flat,
homogeneous model for $z = 0.5 - 1.0$, but in the case with 
$H_0^{\rm II}/H_0^{\rm I} = 0.80$ or $< 0.80$, the [$m,z$] relation is 
difficult to explain the observed data. 

In Figure 3 (curves c), the cases with $\Omega_0^{\rm II} = 0.45, 
0.6,$ and
$0.8$ are shown in a model with $\Omega_0^{\rm I} = 0.3, \ H_0^{\rm
II}/H_0^{\rm I} = 0.82, \ z_1 = 0.067, \ \alpha = 1.0, \ 
{\rm and} \ \lambda_0^{\rm I} = \lambda_0^{\rm II} = 0.$
In the case with $\Omega_0^{\rm II} = 0.45$ and $0.6,$ the relation
are found to be consistent with those in the above flat, homogeneous 
model for $z = 0.5 - 1.0$, but in the case with $\Omega_0^{\rm II} =
0.8$ we have less consistency.  

Here let us examine the lensing effect to the [$m,z$] relation.
In Figure 3 (curves d), the cases with $\alpha = 0.0, 0.5,$ and 
$1.0$ are shown
in a model with $(\Omega_0^{\rm I}, \Omega_0^{\rm II}) = (0.3, 0.6), 
\ H_0^{\rm II}/H_0^{\rm I} = 0.82, \ z_1 = 0.067, \ 
{\rm and} \ \lambda_0^{\rm I} = \lambda_0^{\rm II} = 0.$
Compared with the case  $\alpha = 1.0$, the relations for $\alpha =
0.0$ and $0.5,$ give larger magnitudes especially at epochs $z > 1.0$.
For $\alpha = 0.5$, the magnitudes are by about $0.1$ and $0.2$ mag
larger than those for $\alpha = 1.0$ in the relations at epochs $z =
1.0$ and $2.0$, respectively. If the value  $\alpha = 0.5$ is
realistic, the cases with larger $\Omega_0^{\rm II}$ and smaller
$H_0^{\rm II}/H_0^{\rm I}$ may be consistent with the observed data.

Next we consider the cases with a nonzero cosmological constant.
In Figure 4 (curves a), the cases with $\lambda_0^{\rm II} = 0.0$ 
and $0.4$ are shown in a model with $(\Omega_0^{\rm I}, 
\Omega_0^{\rm II}) = (0.3, 0.6), 
\ H_0^{\rm II}/H_0^{\rm I} = 0.82, \ z_1 = 0.067, \ 
{\rm and} \ \alpha = 1.0$. In the case with $\lambda_0^{\rm II} =
0.4$, the space in the outer region is spatially flat. In this case
the role of the cosmological constant to the accelerating behavior is
not dominant, but supplementary to the role of the local void.

In Figure 4 (curves b), the cases with $H_0^{\rm II}/H_0^{\rm I} 
= 0.80, 0.82,$
and $0.87$ are shown in a model with  $(\Omega_0^{\rm I}, 
\Omega_0^{\rm II}) = (0.3, 0.6), \ z_1 = 0.067, \ \alpha = 1.0, \ 
{\rm and} \ \lambda_0^{\rm II} = 0.4.$  
In this outer-flat case also, larger $H_0^{\rm II}/H_0^{\rm I}$ gives  
larger magnitudes in the [$m,z$] relation.

Finally the lensing effect in homogeneous models is examined for
comparison.  In Figure 4 (curves c), the cases with  $\alpha = 
0.0, 0.5,$ and
$1.0$ are shown in homogeneous models with  $(\Omega_0, \lambda_0) =
(0.3, 0.7)$ and $(0.3, 0.7)$ for $H_0 = 71$. As in Figure 6, the 
magnitudes in the
relation for $\alpha = 0.5$ are by about $0.1$ and $0.2$ mag larger 
than those for $\alpha = 1.0$ in the relations at epochs $z =
1.0$ and $2.0$, respectively.

\section{Concluding remarks}

As for the [$m, z$] relation in cosmological models with a local void,
we studied the parameter dependence of their accelerating behavior,
and found that the local void with $r_1 \la 200 h^{-1}$ Mpc,
$H_0^{\rm II}/H_0^{\rm I} \ga 0.82$, and $\Omega_0^{\rm II} \la 0.6$ 
is appropriate for explaining the accelerating behavior of SNIa without 
cosmological constant, that the lensing with $\alpha \sim 0.5$ is
effective at epochs of $z \ga 1.0$, and that the cosmological
constant ($\lambda_0^{\rm II} \sim 0.4$) necessary for flatness in the outer
region has a role supplementary to the accelerating behavior.
On the basis of these results we can determine in the next step what 
values of the model parameters are best in the direct comparison with
 the observational data of SNIa.

In the Universe with CDM matter, the probability that the
inhomogeneity of Hubble constant $\delta H/H \sim 0.2$ on scales $\sim
200$ Mpc associated with general density perturbations is realized is 
extremely small, as was clarified and discussed by Turner et al. 
(1992), Nakamura and Suto (1995), Shi and Turner (1998), and Wang et
al. (1998).  The constraint from CMB dipole anisotropy was also
discussed by Wang et al.(1998). It should be noticed here that the
{\it spherical} void which we are considering is exceptionally 
compatible with the constraint from CMB dipole anisotropy, in 
spite of the above
large deviation of Hubble constant, as long as the observers are 
near the centre (cf. Paper 1). Inversely it may be suggested that the
local void on scales $\sim 200$ Mpc must be {\it spherical} or 
{\it nearly spherical}, if its existence is real.

In comparison with observations of the galactic number count -
magnitude relation, on the
other hand, Phillips and Turner (1998) have once studied the possibility
of an underdense region on scales of $\sim 300 h^{-1}$ Mpc. However,
a necessary wall for the mass compensation has not been considered in 
their simple models and, in the 
small-angle observations of the above relation
the boundary between the inside region and the outside region 
was rather vague in contrast to the large-angle redshift surveys.

In the near future the void structure on scales of $\sim 200 h^{-1}$
Mpc will be clarified by the galactic redshift survey of SDSS in the
dominant part of whole sky. Then, observational cosmology will be
developed taking into account that we are in a local void.

The author thanks K. Shimasaku for discussions on galactic redshift
surveys and V. M$\ddot{\rm u}$ller for helpful comments.
 This work was supported by Grant-in Aid for Scientific Research 
(No.~12440063) from the Ministry of Education, Science, Sports and
Culture, Japan. 


\label{lastpage}


\begin{thebibliography}{99}

\bibitem{bar} Barber, A.~J., 2000, MNRAS, 318, 195
\bibitem{blant} Blanton, M.~R., Dalcanton, J., Eisenstein, J., Loveday, 
J., Strauss, M.~A., SubbaRau, M., Weinberg, D.~H., Anderson, Jr., J.~E.,
et al., astro-ph/0012085
\bibitem{bro} Broadhurst, T.~J., Ellis, R.~S., Koo, D.~C., Szalay, A.~S., 
1990, Nature,343, 726
\bibitem{cole} Cole, S., Norberg, P., Baugh, C.~M., Frenk, C.~S.,
BlandHawthorn, J., Bridges T., Cannon, R., Colless, M., astro-ph/0012429 
\bibitem{dal} Dale, D.~A. Giovanelli, R., Haynes, M.~P., Hardy, E.,
Campusano, L., 1999, ApJ, 510, L11 
\bibitem{dr} Dyer, C.~C., Roeder, R.~C., 1973, ApJ, 180, L31
\bibitem{ein}  Einasto, J., Einasto, M., Gottl$\ddot{\rm o}$ber, S.,
M$\ddot{\rm u}$ller, V., Saar, V., Starobinsky, A.~A., Tago, E., Tucker, D.,
Andernach, H., Frisch, P., 1997, Nature, 385,139
\bibitem{eist} Einasto, J., Einasto, M., Tago, E., Starobinsky, A.~A.,
Atrio-Barandela, F., M$\ddot{\rm u}$ller, V., Knebe, A., Cen, R., 1999, ApJ, 519, 469
\bibitem{folk} Folkes, S., Ronen, S., Price, I., Lahav, O., Colless,
M., Maddox, S., Deeley, K., Glazebrook, K, et al., 1999, MNRAS, 308, 459
\bibitem{gar} Garnavich, P.~M., Kirshner, R.~P., Challis, P., Tonry, J.,
Gilliland, R.~L., Smith, R.~C., Clocchiatti, A., Diercks, A., et al. ,
 1998, ApJ, 493, L53
\bibitem{giov} Giovanelli, R., Haynes, M.~P., Freudling, W., da Costa,
 L., Salzer, J., Wegner, G., 1998, ApJ, 505, L91
\bibitem{giov9} Giovanelli, R., Dale, D.~A., Haynes, M.~P., Hardy, E.,
Campusano, L., 1999, ApJ, 525, 25
\bibitem{good} Goodwin, S.~P., Thomas, P.~A., Barber, A.~J., Gribbin, J, 
Onuora, L.I., 1999, astro-ph/9906187
\bibitem{holz} Holz, D.~E., 1998, ApJ, 506, L1
\bibitem{hud} Hudson, M.~J., Smith, R.~J., Lucey, J.~R., Schlegel,
D.~J., Davies, R.~L., 1999, ApJ, 512, L79
\bibitem{kant} Kantowski, R., 1998, ApJ, 507, 483 
\bibitem{kim} Kim, A.~G., Gabi, S., Goldhaber, G., Groom, D.~E., Hook,
I.~M., Kim, M.~Y., Lee, J.~C., Pennypacker, C.~R., et al. , 1997, ApJ,
476, L63 
\bibitem{lan} Landy, S.~D., Shectman, S.~A., Lin, H., Kirshner, R.~P.,
Oemler, A.~A., Tucker, D., 1996, ApJ, 456, L1
\bibitem{mari} Marinoni, C., Monaco, P., Giuricin, G, Costantini,
B., 1999, ApJ, 521, 50 
\bibitem{marz} Marzke, R.~O., da Costa, L.~N., Pellegrini, P.~S.,
Willmer, C.~N.~A., Geller, M.~J., 1998, ApJ, 503, 617 
\bibitem{naksut} Nakamura, T.~T., Suto, Y., 1995, ApJ, 447, L65
\bibitem{perl} Perlmutter, S., Aldering, G., Goldhaber, G., Knop,
R.~A., Nugent, P., Groom, D.~E., Castro, P.~G., Deustua, S. et al. , 
1999, ApJ, 517, 565
\bibitem{phill} Phillips, L.~A., Turner, E.~L. 1998, astro-ph/9802352
\bibitem{porc} Porciani, C., Madau, P., 2000, ApJ, 532, 679
\bibitem{riessa} Riess, A.~G., Filippenko, A.~V., Challis, P.,
Clocchiatti, A., Diercks, A., Garnavich, P.~M., Gilliland, R.~L., et
al. , 1998, AJ, 116, 1009
\bibitem{riessb} Riess, A.~G., Filippenko, A.~V., Li, W., Schmidt, B.,
 2000, AJ, 118, 2668
\bibitem{riessc} Riess, A.~G., 2000, PASP, 112, 1284
\bibitem{schect} Shectman, S.~A., Landy, S.~D., Oemler, A., Tucker,
D.~L., Lin, H., Kirshner, R.~P., Schecter, P.~L., 1996, ApJ, 470, 172
\bibitem{sch} Schmidt, B.~P., Suntzeff, N.~B., Pillips, M.~M., Schommer,
R.~A., Clocchiatti, A., Kirshner, R.~P., Garnavich, P., Challis, P., et al. ,
 1998, ApJ, 507, 46
\bibitem{sef} Schneider, P., Ehler, J., Falco, E.~E., 1992,
Gravitational Lenses (Berlin: Springer)
\bibitem{shi} Shi, X.,  Turner, M.~S., 1998, ApJ, 493, 519
\bibitem{tomd} Tomita, K., 1998, Prog. Theor. Phys. 100, 79; \
astro-ph/9806047 
\bibitem{tomc} Tomita, K., 1999, Prog. Theor. Phys. Suppl. 133, 155; \
astro-ph/9904351
\bibitem{tma} Tomita, K., 2000a, ApJ, 529, 26; \ astro-ph/9905278
\bibitem{tmb} Tomita, K., 2000b, ApJ, 529, 38; \ astro-ph/9906027
\bibitem{tmc} Tomita, K., 2001, Prog. Theor. Phys., 105, No.3; \ 
astro-ph/0005031
\bibitem{tco} Turner, E.~L., Cen, R., Ostriker, J.~P., 1992, AJ, 103,1427
\bibitem{wang} Wang, Y., Spergel, D.~S., Turner, E.~L., 1998, ApJ, 489, 1
\bibitem{will} Willick, J.~A., 1999, ApJ, 522, 647
\bibitem{zucc} Zucca, E., Zamorani, G., Vettolani, G., Cappi, A.,
Merighi, R., Mignoli, M., MacGillivray, H., Collins, C. et al. ,
1997, A\&A, 326, 477


\end{thebibliography}
\end{document}